\documentclass[%
aip,
 amsmath,amssymb,
 reprint,%
]{revtex4-2}

\usepackage{graphicx}
\usepackage{dcolumn}
\usepackage{bm}

\usepackage[utf8]{inputenc}
\usepackage[T1]{fontenc}
\usepackage{mathptmx}
\usepackage{etoolbox}
\usepackage{amsmath,array,graphicx}
\usepackage{kantlipsum}

\usepackage{hyperref, soul}
\usepackage{graphicx,color,xcolor}

\usepackage{lineno}

\makeatletter
\def\@email#1#2{%
 \endgroup
 \patchcmd{\titleblock@produce}
  {\frontmatter@RRAPformat}
  {\frontmatter@RRAPformat{\produce@RRAP{*#1\href{mailto:#2}{#2}}}\frontmatter@RRAPformat}
  {}{}
}%
\makeatother
\begin{document}

\preprint{AIP/123-QED}

\title[]{Systematic Improvements in Transmon Qubit Coherence Enabled by Niobium Surface Encapsulation}
\author{Mustafa Bal}
\thanks{These authors contributed equally}
\affiliation{Superconducting Quantum Materials and Systems Division, Fermi National Accelerator Laboratory (FNAL), Batavia, IL 60510, USA}
\author{Akshay A. Murthy}
\thanks{These authors contributed equally}
\affiliation{Superconducting Quantum Materials and Systems Division, Fermi National Accelerator Laboratory (FNAL), Batavia, IL 60510, USA}
\author{Shaojiang Zhu}
\thanks{These authors contributed equally}
\affiliation{Superconducting Quantum Materials and Systems Division, Fermi National Accelerator Laboratory (FNAL), Batavia, IL 60510, USA}
\author{Francesco Crisa}
\thanks{These authors contributed equally}
\affiliation{Superconducting Quantum Materials and Systems Division, Fermi National Accelerator Laboratory (FNAL), Batavia, IL 60510, USA}
\author{Xinyuan You}
\affiliation{Superconducting Quantum Materials and Systems Division, Fermi National Accelerator Laboratory (FNAL), Batavia, IL 60510, USA}
\author{Ziwen Huang}
\affiliation{Superconducting Quantum Materials and Systems Division, Fermi National Accelerator Laboratory (FNAL), Batavia, IL 60510, USA}
\author{Tanay Roy}
\affiliation{Superconducting Quantum Materials and Systems Division, Fermi National Accelerator Laboratory (FNAL), Batavia, IL 60510, USA}
\author{Jaeyel Lee}
\affiliation{Superconducting Quantum Materials and Systems Division, Fermi National Accelerator Laboratory (FNAL), Batavia, IL 60510, USA}
\author{David van Zanten}
\affiliation{Superconducting Quantum Materials and Systems Division, Fermi National Accelerator Laboratory (FNAL), Batavia, IL 60510, USA}
\author{Roman Pilipenko}
\affiliation{Superconducting Quantum Materials and Systems Division, Fermi National Accelerator Laboratory (FNAL), Batavia, IL 60510, USA}
\author{Ivan Nekrashevich}
\affiliation{Superconducting Quantum Materials and Systems Division, Fermi National Accelerator Laboratory (FNAL), Batavia, IL 60510, USA}
\author{Andrei Lunin}
\affiliation{Superconducting Quantum Materials and Systems Division, Fermi National Accelerator Laboratory (FNAL), Batavia, IL 60510, USA}
\author{Daniel Bafia}
\affiliation{Superconducting Quantum Materials and Systems Division, Fermi National Accelerator Laboratory (FNAL), Batavia, IL 60510, USA}
\author{Yulia Krasnikova}
\affiliation{Superconducting Quantum Materials and Systems Division, Fermi National Accelerator Laboratory (FNAL), Batavia, IL 60510, USA}

\author{Cameron J. Kopas}
\affiliation{Rigetti Computing, Berkeley, CA 94710, USA}%
\author{Ella O. Lachman}
\affiliation{Rigetti Computing, Berkeley, CA 94710, USA}%
\author{Duncan Miller}
\affiliation{Rigetti Computing, Berkeley, CA 94710, USA}%
\author{Josh Y. Mutus}
\affiliation{Rigetti Computing, Berkeley, CA 94710, USA}%
\author{Matthew J. Reagor}
\affiliation{Rigetti Computing, Berkeley, CA 94710, USA}%
\author{Hilal Cansizoglu}
\affiliation{Rigetti Computing, Berkeley, CA 94710, USA}%
\author{Jayss Marshall}
\affiliation{Rigetti Computing, Berkeley, CA 94710, USA}%
\author{David P. Pappas}
\affiliation{Rigetti Computing, Berkeley, CA 94710, USA}%
\author{Kim Vu}
\affiliation{Rigetti Computing, Berkeley, CA 94710, USA}%
\author{Kameshwar Yadavalli}
\affiliation{Rigetti Computing, Berkeley, CA 94710, USA}%

\author{Jin-Su Oh}
\affiliation{Ames Laboratory, U.S. Department of Energy, Ames, IA 50011, USA}
\author{Lin Zhou}
\affiliation{Ames Laboratory, U.S. Department of Energy, Ames, IA 50011, USA}
\author{Matthew J. Kramer}
\affiliation{Ames Laboratory, U.S. Department of Energy, Ames, IA 50011, USA}

\author{Florent Q. Lecocq}
\affiliation{National Institute of Standards and Technology, Boulder, CO, USA}

\author{Dominic P. Goronzy}
\affiliation{Department of Materials Science and Engineering, Northwestern University, Evanston, IL, 60208, USA}
\author{Carlos G. Torres-Castanedo}
\affiliation{Department of Materials Science and Engineering, Northwestern University, Evanston, IL, 60208, USA}
\author{Graham Pritchard}
\affiliation{Department of Materials Science and Engineering, Northwestern University, Evanston, IL, 60208, USA}
\author{Vinayak P. Dravid}
\affiliation{Department of Materials Science and Engineering, Northwestern University, Evanston, IL, 60208, USA}
\affiliation{The NU\textit{ANCE} Center, Northwestern University, Evanston, IL, 60208, USA}
\affiliation{International Institute of Nanotechnology, Northwestern University, Evanston, IL, 60208, USA}
\author{James M. Rondinelli}
\affiliation{Department of Materials Science and Engineering, Northwestern University, Evanston, IL, 60208, USA}
\author{Michael J. Bedzyk}
\affiliation{Department of Materials Science and Engineering, Northwestern University, Evanston, IL, 60208, USA}
\author{Mark C. Hersam}
\affiliation{Department of Materials Science and Engineering, Northwestern University, Evanston, IL, 60208, USA}
\affiliation{Department of Chemistry, Northwestern University, Evanston, IL 60208, USA}
\affiliation{Department of Electrical and Computer Engineering, Northwestern University, Evanston, IL 60208, USA}
\author{John Zasadzinski}
\affiliation{Department of Physics, Illinois Institute of Technology, Chicago, IL, 60616, USA}
\author{Jens Koch}
\affiliation{Department of Physics and Astronomy, Northwestern University, Evanston, IL 60208, USA}
\affiliation{Center for Applied Physics and Superconducting Technologies, Northwestern University, Evanston, IL 60208, USA}
\author{James A. Sauls}
\affiliation{Hearne Institute of Theoretical Physics, Department of Physics and Astronomy, Louisiana State University, Baton Rouge, LA 70803, USA}
\author{Alexander Romanenko$^*$}
\affiliation{Superconducting Quantum Materials and Systems Division, Fermi National Accelerator Laboratory (FNAL), Batavia, IL 60510, USA}
\author{Anna Grassellino$^*$}
\affiliation{Superconducting Quantum Materials and Systems Division, Fermi National Accelerator Laboratory (FNAL), Batavia, IL 60510, USA}
\email{Corresponding authors: aroman@fnal.gov, annag@fnal.gov}

\date{\today}

\begin{abstract}

We present a novel transmon qubit fabrication technique that yields systematic improvements in T$_1$ relaxation times. We fabricate devices using an encapsulation strategy that involves passivating the surface of niobium and thereby preventing the formation of its lossy surface oxide. By maintaining the same superconducting metal and only varying the surface structure, this comparative investigation examining different capping materials, such as tantalum, aluminum, titanium nitride, and gold, and film substrates across different qubit foundries definitively demonstrates the detrimental impact that niobium oxides have on the coherence times of superconducting qubits, compared to native oxides of tantalum, aluminum or titanium nitride. Our surface-encapsulated niobium qubit devices exhibit T$_1$ relaxation times 2 to 5 times longer than baseline niobium qubit devices with native niobium oxides. When capping niobium with tantalum, we obtain median qubit lifetimes above 300 microseconds, with maximum values up to 600 microseconds, that represent the highest lifetimes to date for superconducting qubits prepared on both sapphire and silicon. Our comparative structural and chemical analysis suggests why amorphous niobium oxides may induce higher losses compared to other amorphous oxides. These results are in line with high-accuracy measurements of the niobium oxide loss tangent obtained with ultra-high Q superconducting radiofrequency (SRF) cavities. This new surface encapsulation strategy enables even further reduction of dielectric losses via passivation with ambient-stable materials, while preserving fabrication and scalable manufacturability thanks to the compatibility with silicon processes.

\end{abstract}

\maketitle

With massive improvements in device coherence times and gate fidelity over the past two decades, superconducting qubits have emerged as a leading technology platform for quantum computing\cite{doi:10.1146/annurev-conmatphys-031119-050605, Wendin_2017, deLeoneabb2823}. Although many of these improvements have been driven through optimized device designs and geometries, the presence of defects and impurities at the interfaces and surfaces in the constituent materials continues to limit performance and serve as a critical barrier in achieving scalable quantum systems.\cite{RN6,PhysRevLett.123.190502, 8993458} Specifically, these uncontrolled defect sites can serve as sources of loss by introducing two-level systems (TLS) or nonequilibrium quasiparticles \cite{RN23,simmonds2004decoherence, Muller_2019, mcdermott2009materials}. As a result, researchers have recently begun to take a materials-oriented approach to understand and eliminate these sources of quantum decoherence in superconducting qubit devices.

Niobium (Nb) has been widely employed as the primary material in superconducting qubits as it possesses the largest critical temperature and superconducting gap of elemental superconductors, making thermal quasiparticle contribution to losses negligible at typical operating temperatures of $\lesssim 50$~mK. It is also highly compatible with industrial-scale processes \cite{6967749}. Furthermore, the Fermilab superconducting radio-frequency (SRF) research group has demonstrated in prior detailed studies of 3D cavities in the quantum regime that devices processed from Nb can sustain photon lifetimes as high as 2 seconds when the surface niobium oxide hosting sources of TLS is removed. \cite{PhysRevApplied.13.034032}. This is $\sim$3 orders of magnitude longer than coherence times reported in the highest-performing transmon qubits \cite{RN24, RN34}, making bare niobium metal an attractive base material for further improvements in 2D superconducting qubits.

These previous measurements have unambiguously identified the surface oxide that forms spontaneously on Nb under ambient conditions as the major source of microwave loss \cite{PhysRevApplied.13.034032, PhysRevLett.119.264801, RN33, Niepce_2020,  https://doi.org/10.48550/arxiv.2108.13352, Burnett_2016, RN11}. Through 3D cavity measurements, we find that the loss tangent of this 5 nm thick oxide is $\sim$~0.1, which is orders of magnitude larger than the losses at the metal/substrate interface as well as those in the underlying substrate \cite{doi:10.1063/5.0017378, PhysRevApplied.18.034013}. As a result, the removal of this oxide has been shown to boost the photon lifetime by 50-200$\times$ in 3D Nb SRF cavities in the TLS-dominated, $<$1K regime. Other studies on 2D devices have since further confirmed the detrimental effect of this oxide \cite{PhysRevApplied.16.014018, PRXQuantum.3.020312}.

Several recent studies have sought to mitigate losses associated with this region. Unfortunately, most methods for avoiding these losses are incompatible with integration into complex or large-scale manufacturing process flows. In one of the successful approaches, the surface oxide was removed by annealing the sample at temperatures at or exceeding 300$^{\circ}$C \cite{PhysRevApplied.13.034032}, with an almost complete elimination of the TLS-induced losses. While this thermal dissolution method is effective, sustained vacuum is required afterwards to prevent the regrowth of the surface oxide when the cavity or qubit is removed from a ultra-high vacuum (UHV) environment. An alternative oxide removal method involving HF as a wet etchant has been explored as well \cite{PRXQuantum.3.020312, PhysRevApplied.16.014018}. This process cannot be performed in vacuum, leading to rapid oxide re-growth afterwards. Furthermore, it can lead to hydrogen incorporation in the underlying niobium which can introduce resistive niobium hydrides \cite{doi:10.1063/1.1597364, Romanenko_2013, lee2021discovery}. Finally, nitrogen plasma passivation techniques have been identified as effective methods to partially suppress oxide formation \cite{10.1063/5.0082755, doi:10.1080/21663831.2022.2126737}.

Here, we propose a new strategy based on surface encapsulation to eliminate and prevent the formation of this lossy Nb surface oxide upon exposure to air. The first method involves depositing \textit{in situ} metal capping layers of Al and Ta onto the Nb films in UHV. The second method involves atomic layer deposition to reduce the native Nb surface oxide by reacting it with a precursor then depositing a thin metal film (TiN) that exhibits a reduced microwave loss. The third method involves milling away the oxide with Ar$^+$ ions and depositing a thin metal layer of Au with e-beam evaporation. Based on our systematic study, we observe that each of these surface passivation strategies effectively eliminate Nb$_2$O$_5$ and yield a clear improvement in coherence times. Of these capping approaches, we find that the Ta capped Nb films exhibit the largest improvement and lead to devices with median relaxation times of 300 $\mu$s. Finally, we explore the scalability of such an approach by repeating fabrication of test devices with the Ta capping strategy at a commercial qubit fabrication and measurement facility and are able to replicate the results on Si substrates.

\section{Experimental Results}

\begin{figure*}
\includegraphics[width= 7in]{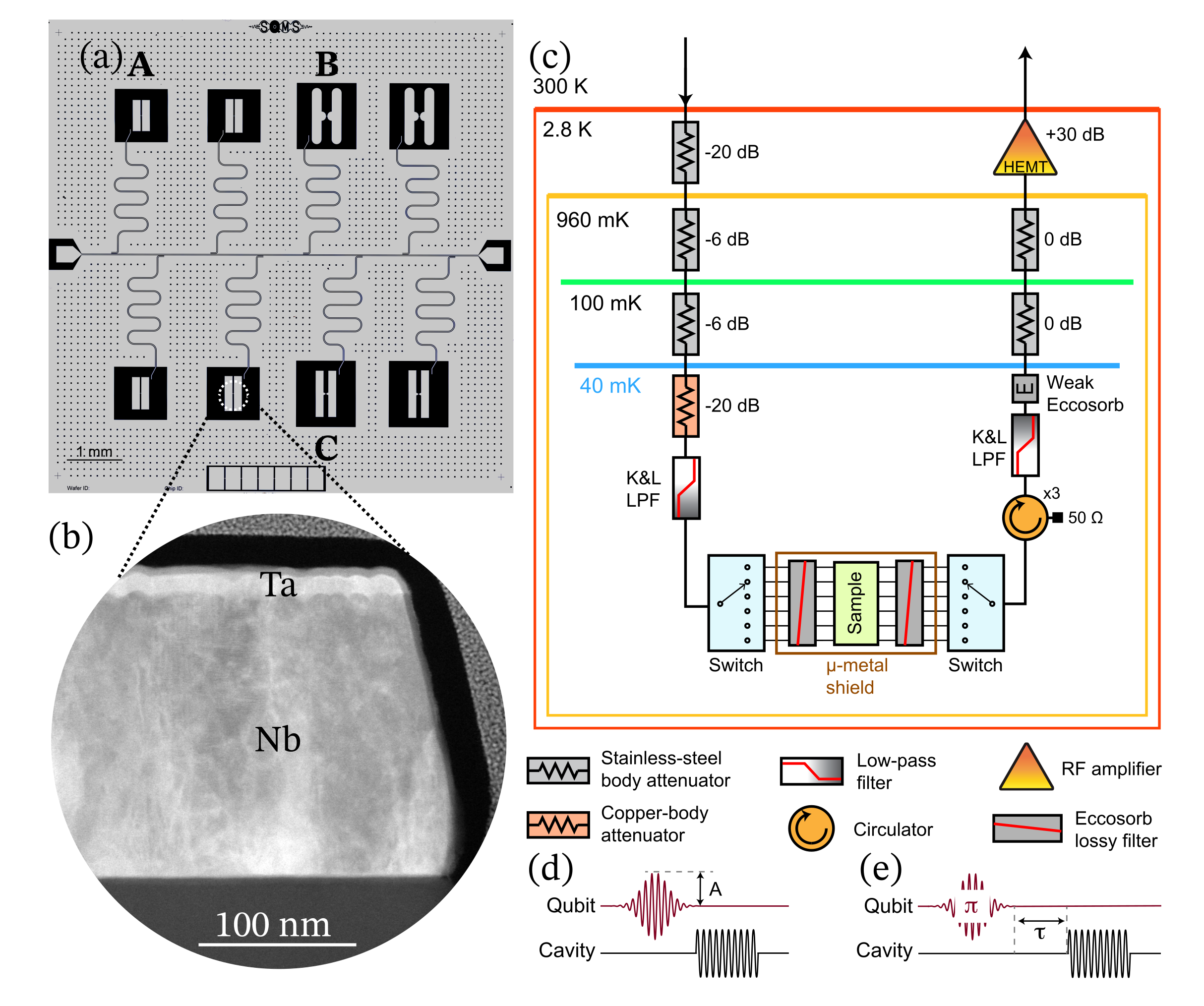}
\caption{(a) 8 qubit chip layout consisting of 3 different geometries. (b) Low magnification annular dark field scanning transmission electron microscopy (ADF-STEM) image taken from a cross-section of a Nb transmon qubit where the Nb film is capped with a Ta metal layer. (c) Cryogenic wiring diagram. (d-e) Pulse scheme for characterization and plots. (d) Pulse for Rabi experiment used to calibrate $\pi$ pulses. (e) Pulse for T$_1$ experiment.}
\label{fig:setup}
\end{figure*}

\begin{figure*}
\includegraphics[width=7in]{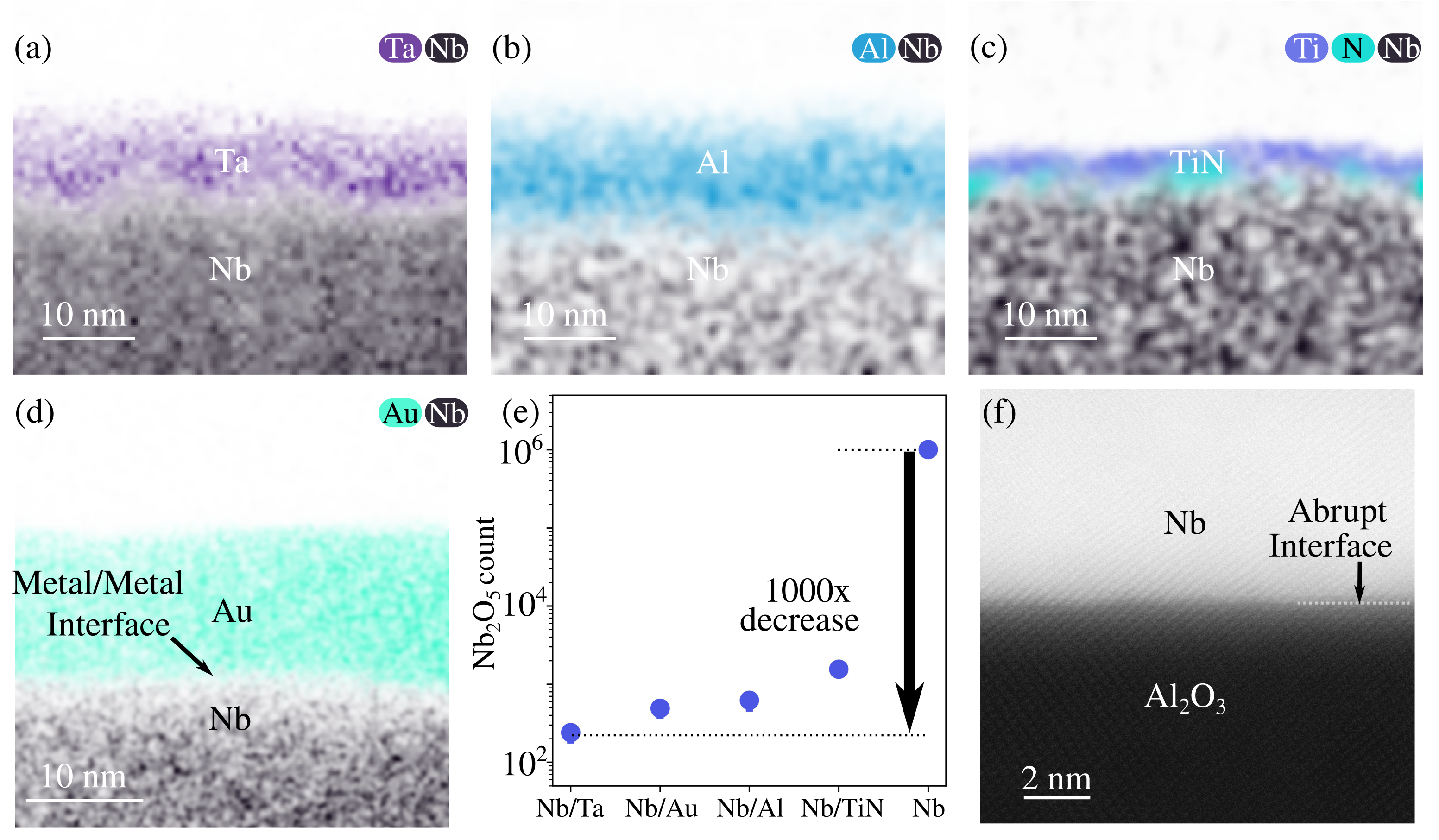}
\caption{Structural and chemical characterization of transmon qubit. (a-d) Chemical phase maps generated through STEM energy dispersive spectroscopy of the Nb films capped with Ta, Al, TiN, and Au, respectively. The Ta, Al, and Au capping layers are roughly 10 nm thick and the TiN layer is roughly 5 nm thick. (e) Plot depicting Nb$_2$O$_5$ counts captured with ToF-SIMS. Each of the capping strategies are effective in mitigating Nb$_2$O$_5$ formation. The Ta capping is particularly effective as a 1000$\times$ decrease in Nb$_2$O$_5$$^-$ counts is observed with this strategy. (f) ADF-STEM image of the metal/substrate interface. Minimal intermixing is observed between Nb and the underlying c-plane sapphire substrate.}
\label{fig:characterization}
\end{figure*}

We fabricated seven sets of qubits for this study as outlined in Table I with device geometries provided in \ref{fig:setup}a. In terms of surface participation ratio, this is largest for device geometry A, followed by device geometry C, and smallest for device geometry B. Details of the fabrication procedure are provided in the Supplementary Information. Results from structural and chemical analysis of the fabricated qubits are provided in Figs. \ref{fig:setup}b and \ref{fig:characterization}. Fig. \ref{fig:setup}b shows a low magnification scanning transmission electron microscopy (STEM) image of a cross-section taken from a superconducting circuit. The superconducting metal (Nb) and metal capping layer (Ta) are labeled. Chemical phase maps generated using STEM energy dispersive spectroscopy (EDS) are presented in Fig. \ref{fig:characterization}a-d for the Nb films capped with different layers. We find that the capping layers are between 10-15 nm, as targeted. Additionally, the layers are spatially distinct with minimal intermixing present. 

\begin{table}[h]
\begin{tabular}{|c|c|p{2.3cm}|c|c|}

\multicolumn{5}{c}{} \\
\hline
\textbf{Substrate} & \textbf{Film} & \centering \textbf{Surface \linebreak Encapsulation} & \textbf{Foundry} & \textbf{Measurement Site}\\
\hline
Sapphire & Nb & \centering - & PNF & Fermilab\\
\hline
Sapphire & Nb & \centering Ta (Sputtering) & PNF & Fermilab\\
\hline
Sapphire & Nb & \centering Al (Sputtering) & PNF & Fermilab\\
\hline
Sapphire & Nb & \centering TiN (ALD) & PNF & Fermilab\\
\hline
Sapphire & Nb & \centering Au (Evaporation) & PNF & Fermilab\\
\hline
Silicon & Nb & \centering - & Rigetti & Rigetti\\
\hline
Silicon & Nb & \centering Ta (Sputtering) & Rigetti & Rigetti\\
\hline

\end{tabular}
\caption{\label{tab:jj} List of fabricated transmon qubits.}
\end{table}

In order to assess the efficacy of the capping layers in preventing Nb$_2$O$_5$ formation, we analyze each film with time-of-flight secondary ion mass spectrometry (ToF-SIMS). This technique combines high mass resolution and sensitivity to both light and heavy elements with $<$100 nm spatial resolution, and has been employed extensively to identify impurities such as oxides in superconducting qubits \cite{RN33, doi:10.1063/5.0079321}. As ToF-SIMS enables identification of different oxide species based on their mass to charge (M/Z) ratios, we are able to resolve that the surface oxide is primarily composed of Nb$_2$O$_5$ based on the presence of localized signal corresponding to the presence of Nb$_2$O$_5$$^-$ ions in this region. 

To better understand how this oxide is impacted by metal capping, we compare the Nb$_2$O$_5$$^-$ signal counts measured from the surface of the baseline Nb sample to the Nb$_2$O$_5$$^-$ signal counts measured at the interface between the capping layer and the Nb metal in the capped Nb samples as indicated in Fig. \ref{fig:characterization}d. We note that this does not refer to total oxide quantity (as aluminum and tantalum themselves oxidize), but specifically to the reduction of the loss channel of particular interest, Nb$_2$O$_5$. The methodology is described in Fig. \ref{SIMS} and the results are presented in Fig. \ref{fig:characterization}e. We find that all of the capping strategies are highly effective in mitigating Nb$_2$O$_5$ formation. The Ta capping is particularly effective as a 1000$\times$ decrease in measured Nb$_2$O$_5$$^-$ is observed with this strategy. The results also suggest that serial sputter deposition may be slightly more effective at protecting again surface oxidation compared to the other two methods. Finally, we observe the presence of a sharp interface between the Nb film and the underlying sapphire as shown in Fig. \ref{fig:characterization}f. In contrast to Nb films grown on silicon where alloyed regions on the order of 5nm have been observed, in Nb/sapphire structures we observe minimal intermixing present at the metal/substrate interface \cite{PhysRevMaterials.6.064402}. 

\begin{figure*}
\includegraphics[width=7in]{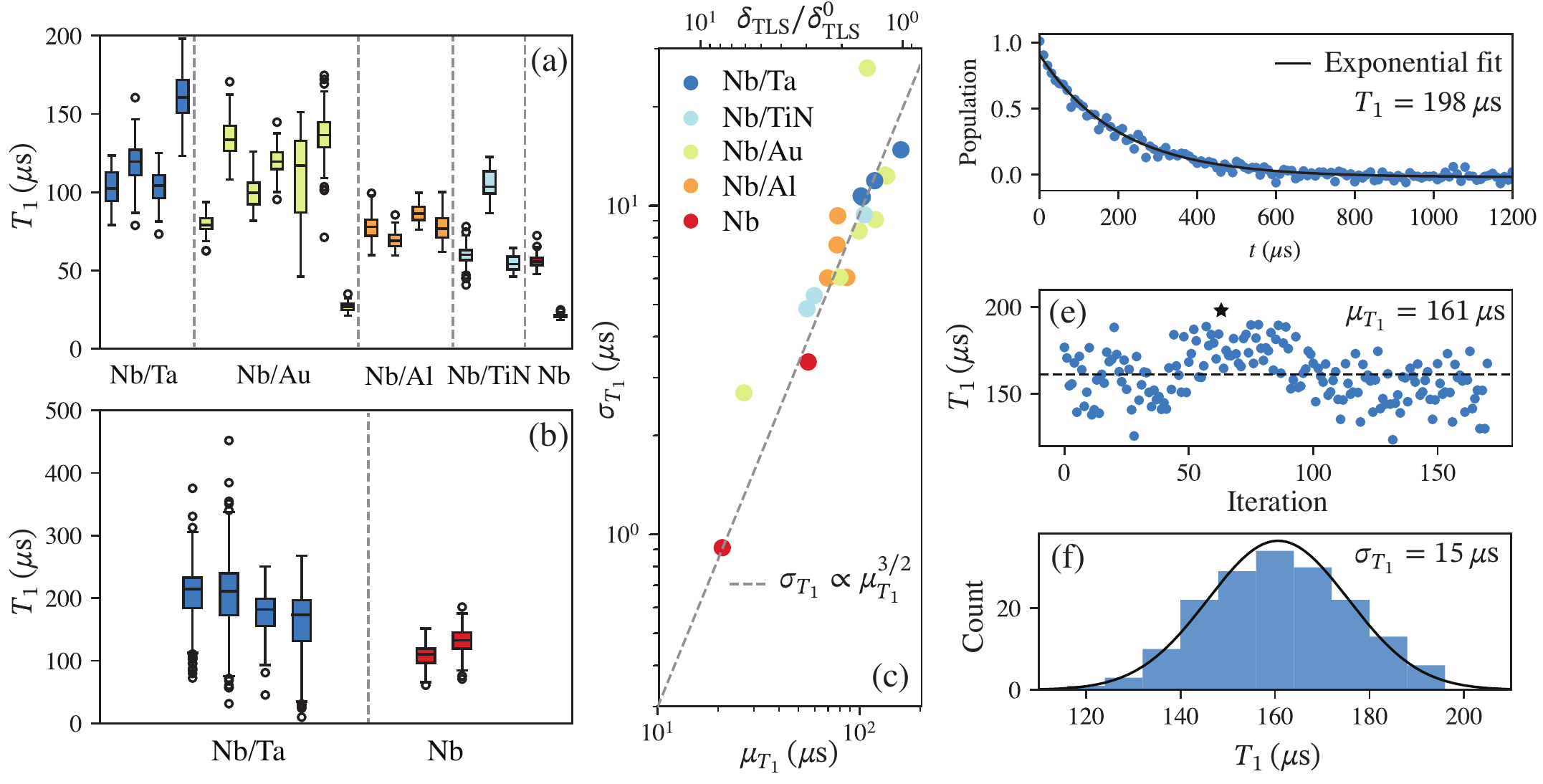}
\caption{Qubit measurement data (a) T\textsubscript{1} comparison of the five sets of qubit devices that were prepared on sapphire substrates. All four Nb/Ta qubits on the chip show T\textsubscript{1}>100$~\mu$s, and the largest T\textsubscript{1} measured for Nb control qubits is $\sim 50~\mu$s. Boxes mark the 25th percentile and the 75th percentile of the measurement distribution over the course of 10 hours of consecutive measurements. The line inside each box represents the median value, and whiskers or circles represent outliers. (b) Measured T\textsubscript{1} values for test devices fabricated on silicon substrates. We observe a clear improvement in terms of the median T\textsubscript{1} value following surface capping of Nb with Ta. (c) Dependence of T\textsubscript{1} standard deviation, $\sigma_{T\textsubscript{1}}$, on the average T\textsubscript{1}, $\mu_{T\textsubscript{1}}$. Different colors correspond to the different encapsulation groups shown in (a). Dashed line shows the best fitting of $\sigma_{T\textsubscript{1}} \propto \mu_{T\textsubscript{1}}^{3/2}$, according to Ref[~\onlinecite{you2022stabilizing}]. It also shows that the Nb/Ta qubit has 5-10 times improvement of TLS loss compared with the Nb qubit, after converting the number of TLS into the tangent loss, $\delta_{TLS}$. (d) Best T\textsubscript{1}=198 $\mu s$. (e) Statistics for T\textsubscript{1} consecutively measured over 10 hours, the average T\textsubscript{1} ($\mu_{T\textsubscript{1}}$) is $161 \mu s$ and the standard deviation ($\sigma_{T\textsubscript{1}}$) is 15$\mu s$. The star shows the iteration that yielded the best T\textsubscript{1} (see (d) for the decay curve). (f) Histogram of the T\textsubscript{1} values in (b), with a Gaussian fit.}
\label{fig:measurements}
\end{figure*}

To most easily observe the impact of the metal capping layer on the superconducting qubit coherence, we performed measurements using the qubit geometry with the largest surface participation ratio (geometry A). These measurements were performed on the capped devices and on baseline Nb devices that were not capped.

The qubit devices are measured inside a dilution refrigerator at a temperature of around 40 mK via dispersive readout. The cryogenic wiring diagram is shown in Fig.~\ref{fig:setup}c. Both qubit and readout pulses are sent through a single RF line. At the mixing chamber plate, two six-pole-single-throw microwave switches are used to direct the signals to the relevant sample and to extract the outgoing signal. After initial qubit spectroscopy a Rabi measurement (Fig.~\ref{fig:setup}d is performed to determine the $\pi$-pulse length followed by T\textsubscript{1} measurements (Fig.~\ref{fig:setup}e).

Because the qubit energy relaxation time, $T_1$, is largely dependent on the material losses in the qubit device, while the dephasing time, $T_2$, is heavily impacted by many environmental factors such as the thermal noise, IR radiation, and cosmic rays \cite{PhysRevA.75.042302, PhysRevA.106.042605}, we focus on qubit T\textsubscript{1} characterization. This qubit T\textsubscript{1} characterization is widely believed to directly reflect the loss due to TLS, which is the focus of this work. The measurement follows the standard procedure \cite{krantz2019quantum}, \textit{i.e.}, the qubit is driven from the ground to the first-excited state by a calibrated $\pi$ pulse. The qubit state is then read after a variable delay. The relaxation times are fitted to a single-exponential decay to extract T\textsubscript{1}. In order to probe both the typical and exceptional T\textsubscript{1} times, we benchmark the T\textsubscript{1} measurement by continuously collecting data for 10 hours for each qubit, as described in Ref [~\onlinecite{burnett2019decoherence}]. Using this data set, we extract the average, standard deviation, and best T\textsubscript{1} values, which we compare across different devices. 

A comparison of the T\textsubscript{1} measured in different devices is provided in Fig. \ref{fig:measurements}(a) and further summarized in Table \ref{tab:qubits_capping}. The Nb qubits capped with Ta as well as the Nb qubits capped with Au have the highest average T\textsubscript{1} (> 100$\mu s$), while the baseline Nb qubits have the lowest average T\textsubscript{1}. The average T\textsubscript{1} value of Nb qubits capped with Al are slightly higher than that of Nb qubits capped with TiN. The improvement in average T\textsubscript{1} for all of the capped devices suggests that reducing the native Nb$_2$O$_5$ (a strong TLS host) on the surface of the Nb film improves qubit energy relaxation times. Whiskers or circles in the plot indicate the maximum and minimum T\textsubscript{1} values observed during the greater than 10 hours measurement window per qubit [see Fig \ref{fig:measurements}(d) for the best T\textsubscript{1} decay curve]. The fluctuations of T\textsubscript{1} over time [shown in Fig.~\ref{fig:measurements}(e)], and its Gaussian distribution [shown in Fig.~\ref{fig:measurements}(f)] have both been observed in literature~\cite{klimov2018fluctuations, burnett2019decoherence}, and are typically considered a signature of qubit lifetime limited by TLS defects residing in the materials.

To quantitatively compare TLS densities from different encapsulations, we calculate the average ($\mu_{T\textsubscript{1}}$) \textit{vs.} standard deviation ($\sigma_{T\textsubscript{1}}$) T\textsubscript{1} for each of the qubits that were measured continuously for 10 hours. According to TLS theory \cite{you2022stabilizing}, $\mu_{T\textsubscript{1}}$ and $\sigma_{T\textsubscript{1}}$ can be represented by the following relations:
\begin{alignat}{2}
\mu_{T\textsubscript{1}} &= \alpha/ (\omega N), \\
\sigma_{T\textsubscript{1}} &= \beta/( \omega N)^{3/2},
\end{alignat}
where $\alpha$ and $\beta$ are temperature-dependent constants and $\omega$ is the qubit frequency, which is treated as constant in this case as the qubit frequencies all lie within the range of 4 and 6 GHz. $N$ is proportional to the number of TLS present in the qubit. From Eqs. (1) and (2), we find that $\sigma_{T\textsubscript{1}} \propto \mu_{T\textsubscript{1}}^{3/2}$ and the data is in agreement with this relationship (dashed line in Fig. \ref{fig:measurements}(c)). Based on this plot, we conclude that Nb qubits capped with Ta and Nb qubits capped with Au exhibit the smallest number of TLS and the highest $\mu_{T\textsubscript{1}}$,  maximum T\textsubscript{1} and fluctuations $\sigma_{T\textsubscript{1}}$. Conversely, the baseline Nb qubit exhibits the greatest number of TLS and therefore the lowest average T\textsubscript{1}. Assuming TLS loss tangent, $\delta_{TLS}$, is proportional to the number of TLS \cite{Muller_2019}, we plot the relative change of $\delta_{TLS}$ for each set of devices, as shown in Fig. \ref{fig:measurements}(c). We find 5-8$\times$ reduction in loss between the Ta-capped Nb devices and the baseline Nb devices. Additionally, we find that the loss associated with Al-capped Nb qubits is slightly lower than TiN-capped Nb qubits. These results are also in agreement with the qubit T\textsubscript{1} measurements.

To assess the reproducibility, scalability and applicability of this capping method to other substrates, Ta-capped and Nb-only devices (with the same test geometry) were fabricated in the Rigetti Computing quantum integrated circuit foundry (Fremont, California) on high-resistivity Si(100) substrates. More details of this fabrication process can be found in Ref [~\onlinecite{8993458}]. The Ta-capped devices exhibit a systematic improvement in median T\textsubscript{1} ($\sim$200 $\mu$s) compared to the baseline Nb qubits ($\sim$120 $\mu$s). A box plot providing information on the measurements performed on both sets of qubits is provided in Fig. \ref{fig:measurements}b with a maximum measured T$_1$ value of 451 $\mu$s. The consistent improvement in median T\textsubscript{1} for the capped Nb devices for both silicon and sapphire substrates supports a performance limitation imposed by the amorphous Nb$_2$O$_5$ surface oxide in un-capped devices. Meanwhile, the T\textsubscript{2}, T\textsubscript{2 echo}, and T$_{\phi}$ values are provided in Fig. \ref{T2}.

\begin{figure}
\includegraphics[width=\columnwidth]{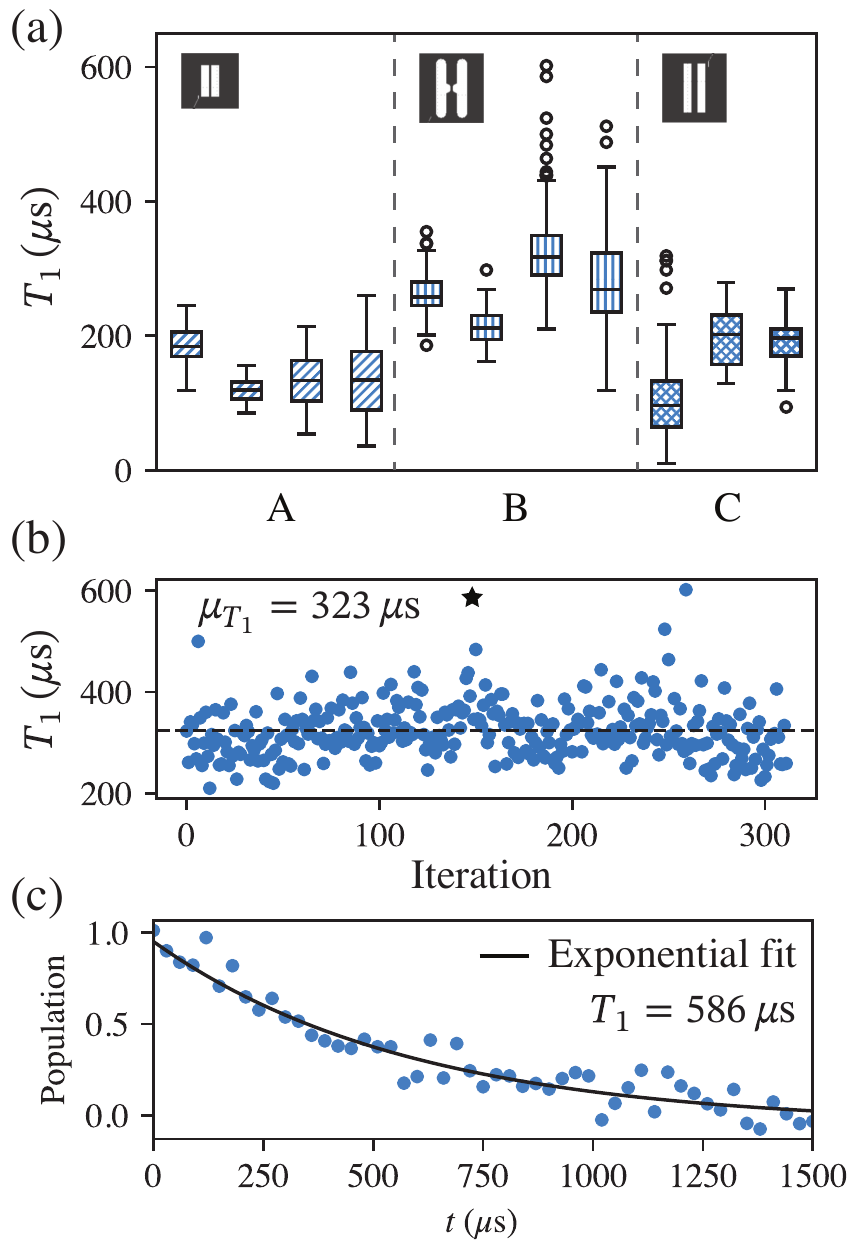}
\caption{Effect of Qubit Geometry (a) T\textsubscript{1} comparison of Nb/Ta qubits as a function of three different qubit geometries prepared on sapphire substrates. Geometry B yields the largest T\textsubscript{1} values. (b) Statistics for T$_1$ consecutively measured over 70 hours for a Geometry B qubit. The average T\textsubscript{1} ($\mu_{T\textsubscript{1}}$) is $323 \mu s$. (c) Decay curve associated with measured T\textsubscript{1} value of 586 $\mu s$ (indicated by star in (b).}
\label{fig:highT1}
\end{figure}

Given these promising T\textsubscript{1} results, we also fabricated larger footprint qubit geometries of the Ta-capped Nb (Geometry B and C) on sapphire that are comparable to those used by groups that have demonstrated the largest T\textsubscript{1} values to date \cite{RN24}. In Figure \ref{fig:highT1}, a comparison of the measured T\textsubscript{1} values for a Ta-capped device is provided for the 3 different geometries that were investigated. These results are also summarized in Table \ref{tab:qubit_geometries}. Geometry B yields the T\textsubscript{1} values with median qubit lifetimes above 300 microseconds, which are in line with the median qubit lifetimes reported by these groups. Further, we observe several individual measurements in excess of 550 $\mu$s that represent the highest lifetimes reported to date. These results, combined with the fact that this approach is reliable with both silicon and sapphire substrates and can be performed at room temperature makes surface encapsulation a very attractive methodology for achieving high coherence qubits. Moreover, future studies will involve gold encapsulation and other similar low-loss capping layers with similar larger footprint geometries to potentially push the envelop of performance even further.

\begin{figure}
\includegraphics[width=\columnwidth]{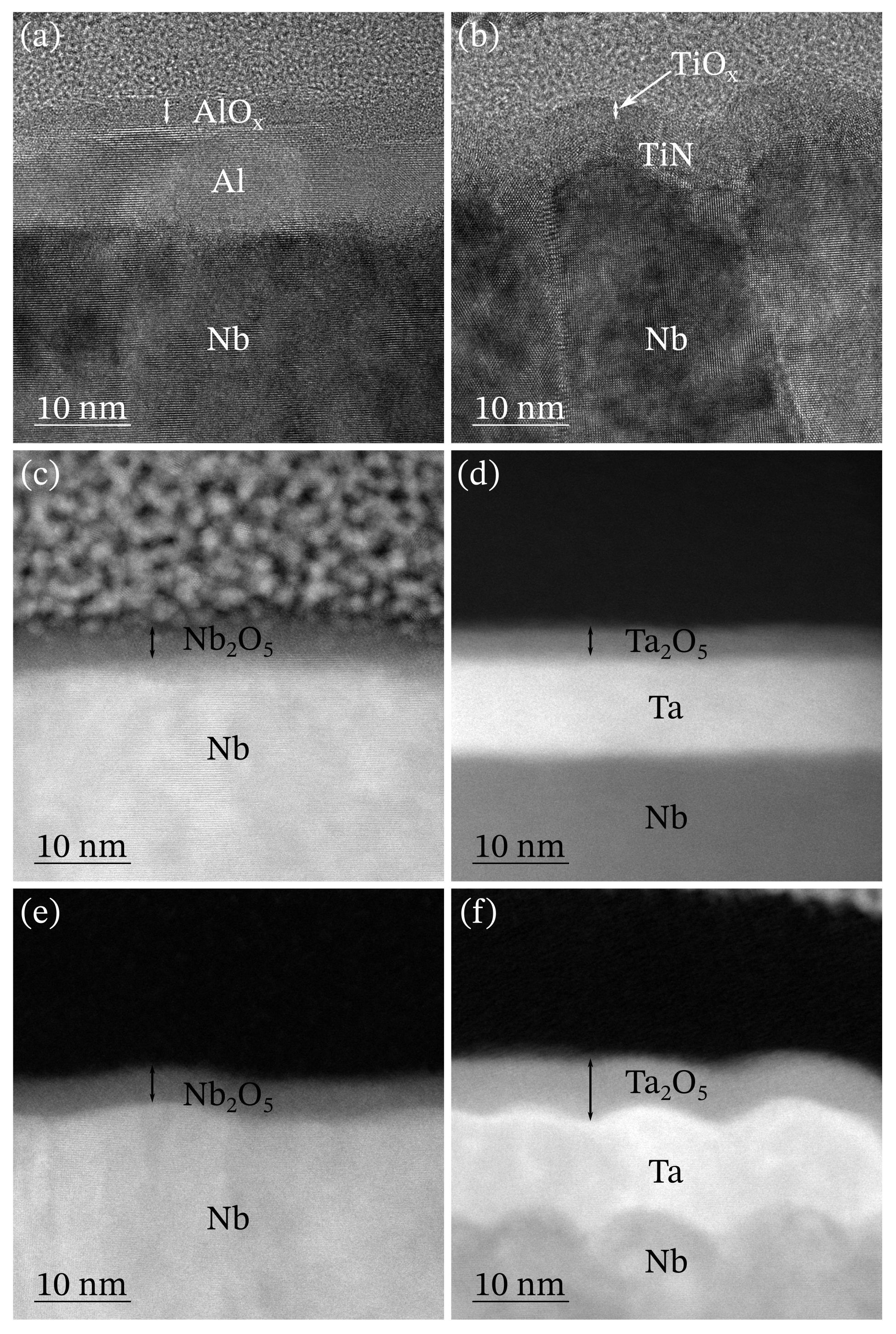}
\caption{Electron microscopy images of surface oxides observed for (a) Al capped Nb, (b) TiN capped Nb, (c) baseline Nb, and (d) Ta capped Nb qubits prepared on sapphire. Similar images of surface oxides for (e) baseline Nb, and (f) Ta capped Nb qubits prepared on silicon are also presented. The oxides are identified with white arrows and tabulated in Table \ref{tab:oxide}.}
\label{fig:oxides}
\end{figure}

Together, these results clearly demonstrate that eliminating Nb$_2$O$_5$ enhances the T$_1$ relaxation time of Nb transmon qubits. In the case of devices capped with Ta, Al, and TiN, we still observe the presence of amorphous oxides of a few nm thickness such as Ta$_2$O$_5$, AlO$_x$, and TiO$_x$, respectively at the sample surface (Fig. \ref{fig:oxides} and Fig. \ref{amorphous_oxides}). By linking modifications solely to the metal/air interface to measured T$_1$ values, we observe a trend where Ta$_2$O$_5$ ranks as the least lossy oxide of those measured, followed by Al$_2$O$_3$, TiO$_2$, and finally, Nb$_2$O$_5$. Further, the loss introduced by these various surface oxides does not appear to be directly correlated to their individual thicknesses as illustrated in Table \ref{tab:oxide}. In particular, 1-2nm TiO$_x$ at the surface of the TiN capped Nb qubit is found to be the thinnest oxide whereas the 5-7 nm Ta$_2$O$_5$ observed at the surface of the Ta capped Nb qubit prepared on silicon is found to be the thickest.

Therefore, our findings help explain previous experimental studies with qubits prepared from Ta metal exhibit improved T$_1$ values \cite{RN24, RN34}. Namely, our results suggest the improved T$_1$  predominantly arises from the presence of a less lossy surface oxide, as opposed to the tantalum film being less lossy compared to niobium.

\begin{table}[h]
\begin{tabular}{|c|c|p{2.3cm}|c|}

\multicolumn{4}{c}{} \\
\hline
\textbf{Substrate} & \textbf{Film} & \centering \textbf{Surface \linebreak Encapsulation} & \textbf{Surface \linebreak Oxide}\\
\hline
Sapphire & Nb & \centering - & 3-5nm Nb$_2$O$_5$\\
\hline
Sapphire & Nb & \centering Ta & 3-5nm Ta$_2$O$_5$\\
\hline
Sapphire & Nb & \centering Al & 2-4nm AlO$_x$\\
\hline
Sapphire & Nb & \centering TiN & 1-2nm TiO$_x$\\
\hline
Sapphire & Nb & \centering Au & Not Observed\\
\hline
Silicon & Nb & \centering - & 3-5nm Nb$_2$O$_5$\\
\hline
Silicon & Nb & \centering Ta & 5-7nm Ta$_2$O$_5$\\
\hline

\end{tabular}
\caption{\label{tab:oxide} Surface oxides observed on fabricated transmon qubits.}
\end{table}

\begin{figure*}
\includegraphics[width=7in]{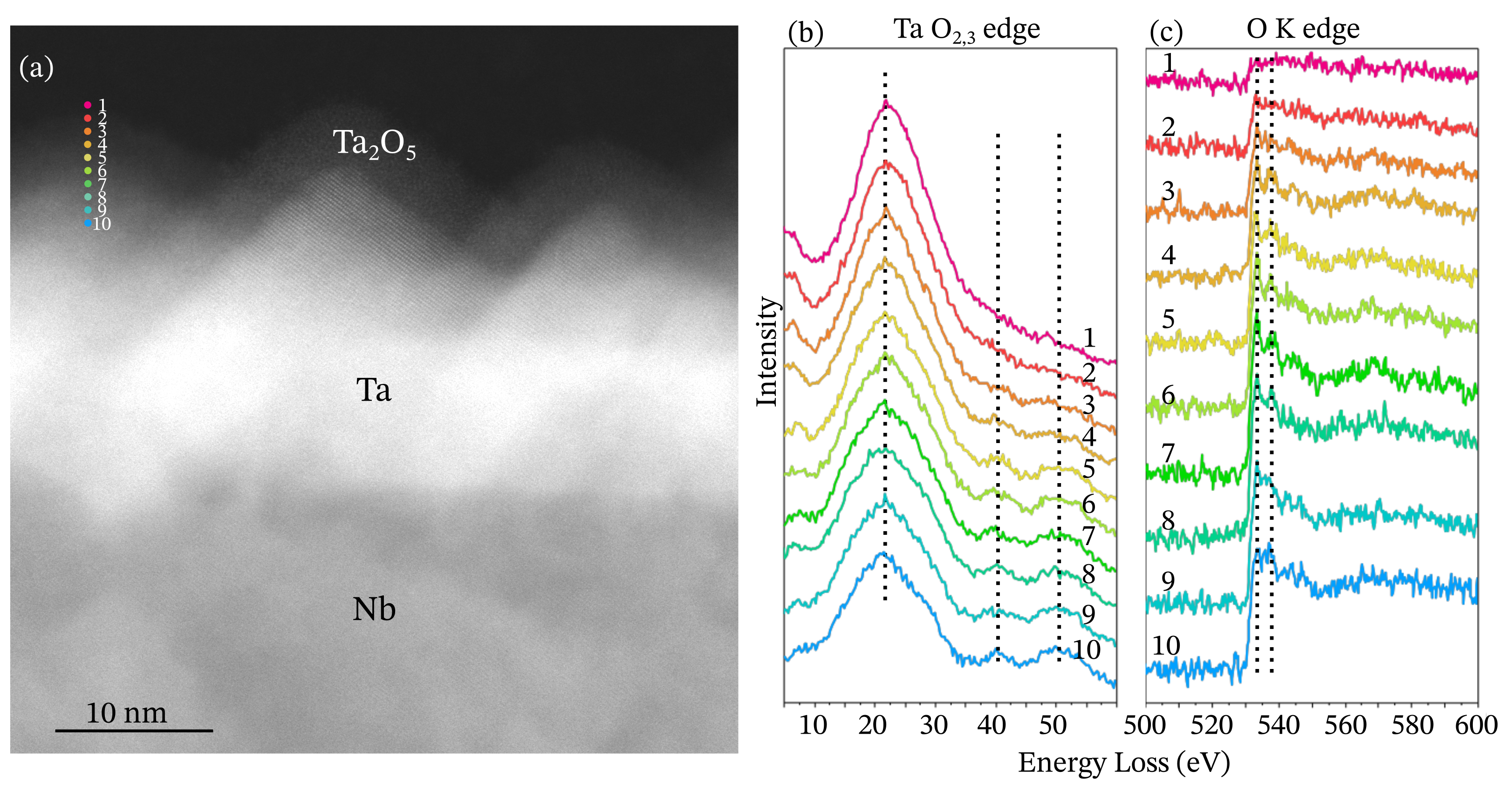}
\caption{(a) Dark field STEM image of the Ta$_2$O$_5$/Ta interface. Electron energy loss spectra (EELS) taken from the locations indicated in (a) are provided in (b) and (c). (b) EELS signal taken from the specified region demonstrating how the tantalum O$_{2,3}$ edge evolves with position. (c) EELS signal taken from the specified region demonstrating how oxygen K edge evolves as a function of position. The dotted lines indicate minimal variation in the position of various features of the spectra. This suggests that the oxygen stoichiometry remains consistent throughout the oxide region.}
\label{fig:EELS}
\end{figure*}

To understand the difference in performance between the capped and baseline samples, we use electron microscopy and, in particular, electron energy loss spectroscopy (EELS) to further evaluate the chemical nature of the Ta oxide. EELS signal captured from points 1-10 on the dark field image of the Ta$_2$O$_5$/Ta interface of the prepared qubit are provided in Fig. \ref{fig:EELS}. From this image, we observe that the oxide thickness of Ta is roughly similar to that observed for Nb oxide (4-5 nm). The Ta oxide is found to be predominantly amorphous based on the presence of diffuse diffraction patterns taken in this region (Fig. \ref{amorphous_oxides_STEM}). Features associated with the tantalum O$_{2,3}$ edge are labeled with a dotted line in Fig. \ref{fig:EELS}b and those associated with the oxygen K edge are labeled with a dotted line in Fig. \ref{fig:EELS}c. We find there are changes in the shape of both set of features, but the position of oxygen K remains constant. This indicates that the Ta predominately exists in a 5+ state. This is in contrast to what has been observed in Nb. For Nb, shifts in the features accompanying the onset of the oxygen K edge are observed as a function of position in the oxide due to changes in the valence state \cite{doi:10.1063/1.3665193, https://doi.org/10.48550/arxiv.2204.06041}. This suggests that the Ta oxide present in these capped samples is largely free of sub-stoichiometric regions. 

Finally, the x-ray reflectivity signal captured from both the Nb sample as well as the Ta-capped Nb sample is provided in Fig. \ref{fig:xrr}. Based on the data fits performed using dynamical scattering theory \cite{https://doi.org/10.1107/S0021889806005073}, we observe that the surface oxide of Nb consists of roughly 4.1 nm of Nb$_2$O$_5$ and 0.5 nm of NbO whereas the surface oxide of Ta consists entirely of 5.9 nm of Ta$_2$O$_5$. This technique suggests that the capped samples are largely free of sub-oxides and is consistent with the TEM findings. 

Theoretical and experimental findings have linked the variable oxygen content in the Nb oxide layer to the presence of local paramagnetic moments. These moments are a source of flux noise, dephasing, and energy loss \cite{sheridan2021microscopic, 5719535}. We hypothesize that the stoichiometric, predominantly Ta$_2$O$_5$ layer of Ta-capped Nb reduces the potential for moment formation compared to the variable oxygen content observed in the Nb oxide layer of uncapped Nb. Additionally, it is possible the Ta oxide layer hosts fewer TLS in the frequency range of the qubit frequency compared to the Nb oxide layer. This is an area of continuing active exploration through experimental and theoretical studies.
In summary, we have implemented different passivation strategies to eliminate and prevent the formation of lossy Nb surface oxide in Nb superconducting qubits. By capping Nb films with Ta, Al, TiN, and Au, we are able to systematically improve the average qubit relaxation times. Of these capping strategies, we find that the Nb film capped with Ta and Nb film capped with Au yield the highest average T\textsubscript{1}. We observe similar improvements in average T\textsubscript{1} when superconducting qubits are prepared on silicon as well as sapphire. Together, this methodology offers a solution to delivering state-of-the-art devices with median T\textsubscript{1} times exceeding 300 $\mu$s that is compatible with industrial-level processes. Further, this method offers a pathway for continuing to suppress the dielectric loss associated with the metal/air interface through further exploration of capping Nb film with ambient-stable layers prior to air exposure and is applicable to the field of superconducting devices broadly (quantum information science, detectors for cosmic science, and particle accelerators). Finally, this study provides definitive proof that niobium oxides exhibit a larger TLS density than tantalum oxide and will help guide future investigations aimed at building a microscopic understanding of TLS sources in superconducting qubits.

\section*{Supplementary Information}
\noindent See supplementary information for wiring diagrams, pulse schemes associated with device measurements, chemical maps, depth profiles and electron diffraction patterns taken from the samples.

\begin{acknowledgments}
\noindent This material is based upon work supported by the U.S. Department of Energy, Office of Science, National Quantum Information Science Research Centers, Superconducting Quantum Materials and Systems Center (SQMS) under contract no. DE-AC02-07CH11359. This work made use of the Pritzker Nanofabrication Facility of the Institute for Molecular Engineering at the University of Chicago, which receives support from Soft and Hybrid Nanotechnology Experimental (SHyNE) Resource (NSF ECCS-2025633), a node of the National Science Foundation’s National Nanotechnology Coordinated Infrastructure. The authors thank members of the SQMS Center for valuable discussions.
\end{acknowledgments}

\section*{Data Availability Statement}

\noindent The data that support the findings of this study are available from the corresponding author upon reasonable request.

\section*{Disclaimer}
Certain commercial equipment, instruments, or materials are identified in this paper in order to specify the experimental procedure adequately. Such identification is not intended to imply recommendation or endorsement by NIST, nor is it intended to imply that the materials or equipment identified are necessarily the best available for the purpose.


\bibliography{aipsamp}

\clearpage
\newpage
\mbox{~}
\clearpage
\newpage
\setcounter{figure}{0}
\setcounter{page}{1}
\setcounter{section}{0}
\setcounter{table}{0}

\renewcommand{\thepage}{S\arabic{page}}
\renewcommand{\thesection}{S\arabic{section}}
\renewcommand{\thetable}{S\arabic{table}}
\renewcommand{\thefigure}{S\arabic{figure}}

\section*{Supplementary Information}
\section*{Methods} \label{Methods}
\subsection*{Qubit Fabrication}
The first four sets of qubits were fabricated on $550~\mu$m thick c-plane sapphire wafers (double-side polished, HEMEX grade HEM Sapphire) at the Pritzker Nanofabrication (PNF) facility. These wafers were first solvent cleaned by sonicating in n-methyl-2-pyrrolidone (NMP) heated to 80$^{\circ}$C followed by sequential soaks in isopropanol, acetone, isopropanol, and deionized (DI) water under mild sonication at room temperature. The sapphire wafers were subject to further cleaning in a mixture of DI water, $30\%$ ammonium hydroxide in water, $30\%$ hydrogen peroxide in water with a 5:1:1 ratio. This cleaning was performed at 65$^{\circ}$C for 5 minutes followed by DI water rinse. Following the surface treatments, the substrate was immediately placed inside an AJA ATC 2200 sputtering system with a base pressure less than $10^{-7}$ Torr. After dehydrating the substrate under vacuum at 200$^{\circ}$C for 30 minutes, the wafers were allowed to cool-down before Nb was sputtered at room temperature. DC magnetron sputtering was performed using a 3-inch diameter Nb target with a metals basis purity of 99.95\% with an Ar flow rate of 30 sccm and partial pressure of 3.5 mTorr. A sputtering power of 600 W was used and the substrate was rotated at 20 rpm. These conditions resulted in a deposition rate of approximately 19 nm/min. For the qubits prepared on silicon substrates the process involved first cleaning the substrate with a RCA surface treatment detailed previously \cite{doi:10.1063/1.3517252} followed by a buffered oxide etch, before immediately loading into the sputter deposition system \cite{8993458}. In this case, Nb films were deposited by high power impulse magnetron sputtering with a 6-inch diameter Nb target having a metal basis purity of 99.9995\% with a base pressure less than 1E-8 Torr at room temperature.

For the Nb samples capped with Ta or Al, the capping layers were immediately sputtered onto the Nb film \textit{in situ}, i.e. without breaking vacuum in the deposition chamber. Al was deposited from a 3-inch diameter Al target (metals basis purity of 99.9995\%) with a DC sputtering power of 500 W. Ta was deposited from a 3-inch diameter Ta target (metals basis purity of 99.95\%) with an RF sputtering power of 600 W. The flow rate and substrate rotation speed were kept constant. The Ar partial pressure was increased to 5 mTorr for the Ta encapsulation layer.

For the Nb films capped with TiN, the sample was removed from the sputter deposition system and immediately loaded into an Ultratech/Cambridge Fiji G2 Plasma-Enhanced ALD system for atomic layer deposition of TiN. The TiN ALD recipe comprised of a 250 ms TDMAT, (Ti(N(CH$_3$)$_2$)$_4$) precursor dose and a 30 s N$_2$ plasma exposure. Each of these pulses was separated by a 5 s purge. This process was repeated for 40 cycles and the sample was held at 270$^{\circ}$C during the course of this deposition. These capped layers were all between 10-15 nm in thickness.

For the Nb films capped with Au, the sample was removed from the sputter deposition system and immediately loaded into an Angstrom EvoVac Electron Beam Evaporator system with a base pressure of 2E-7 Torr. The Nb surface oxide was milled away with Ar$^+$ ions for 15 min with a beam voltage of 400 V, an accelerator voltage of 80 V and an ion current of 15 mA. A 10 nm layer of Au was subsequently deposited using e-beam deposition at a rate of 1 angstrom/second. During this process, the substrate was rotated at 10 rpm.

Following film deposition, the superconducting microwave circuit was patterned with photolithography using a Heidelberg MLA 150 Advanced Maskless Aligner. After exposure and development of the photoresist, the circuit layer was defined by dry etching using an inductively coupled plasma - reactive ion etching (ICP-RIE) system. Dry etching was performed using a mixture of Cl$_2$:BCl$_3$:Ar with flow rates 30:10:10 sccm. The etching process was halted following the etching of the Nb metal as well as some of the underlying substrate. In the case of sapphire, the substrate was overetched by 20nm and in the case of silicon, the substrate was overetched by 70nm. The remaining photoresist was stripped from the patterned substrate first by ashing in O$_2$ (to remove the hardened photoresist on the surface) and then by soaking in NMP heated to 80$^{\circ}$C for several hours. 

The Dolan bridge style Al/AlOx/Al Josephson junctions are fabricated using the well-known double angle shadow evaporation technique and solvent lift off for the samples prepared on sapphire \cite{doi:10.1063/1.89690}. The junction patterns are exposed using an electron beam with an accelerating voltage of 100kV. The JJ layers for the samples prepared on silicon are deposited using a double-angle bridge-less lift-off lithography process. In a subsequent lithography and metal lift-off step, galvanic connection between the Josephson junctions and the microwave circuitry is established. Ar ion milling is employed to remove surface oxides immediately prior to Al deposition. Following lift-off of the junction leads, the surface is exposed to a mild UV ozone treatment for 10 min to remove residual resist.

The geometry of the qubit consists of a pair of rectangular shunting capacitor paddles joined by a single Josephson junction as seen in Fig. \ref{fig:setup}a. The paddle size, hence the capacitance, is varied while the gap is kept fixed. A geometry similar to this has exhibited greater than $300~\mu$s qubit coherence times in the past\cite{gordon2022environmental}. To minimize the influence of the magnetic flux noise, all qubits are designed to have fixed frequencies. Following the circuit quantum electrodynamics (cQED) architecture \cite{blais2021circuit}, each qubit is capacitively coupled to a single quarter-wave readout resonator.  Each of these resonators is inductively coupled to a $50 ~\Omega$ transmission line where a microwave probe tone can be sent to the resonator. On a single 7.5$\times$7.5 mm$^2$ sapphire chip, there are four qubits with this rectangular-paddle geometry. The resonator frequencies are evenly spaced between 6.8 and 7.2 GHz with qubit frequencies ranging from 3.4-6 GHz as summarized in Tables S1 and S2.

\subsection{Qubit Measurement Setup and Method}

Each qubit chip was packaged in a gold plated copper box, where it is mounted on a cold finger directly attached to the dilution refrigerator (DR) mixing plate. Both the enclosure and the cold finger are made of gold-coated copper to reduce thermal resistance between the qubit chip and the enclosure. Each qubit chip is protected from IR photons with Eccosorb filters on both input and output lines. All qubit enclosures and Eccosorb filters are enclosed within magnetic shielding to reduce flux noise and are wired to cryogenic switches anchored on the mixing stage. This way, all qubits share the same microwave input and output lines, which enables comparison of different qubits. A total of 52 dB attenuation is distributed across different temperature stages to effectively suppress thermal noise below $10^{-3}$ noise photon number \cite{krinner2019engineering}. Three cryogenic isolators with a total of 60 dB isolation at the output line reduces the backaction noise arising from the high electron mobility transistor (HEMT) amplifier mounted on the 3K stage.  Low-pass filters (K\&L Microwave) are used to remove the high-frequency noise above 10 GHz.

The qubit state is measured via dispersive readout, in which the qubit frequency is far detuned from the resonator frequency ($\Delta = f_q - f_r$) so that $\Delta \gg \kappa, \textit{g}$, where $\kappa$ and \textit{g} are the resonator linewidth and qubit-resonator coupling strengths, respectively \cite{krantz2019quantum}. In this case, the qubit and resonator do not directly exchange energy. Instead, the qubit induces a state-dependent frequency shift of the resonator. By interrogating the resonator, the qubit state can be inferred.

Qubit measurements of devices on Si substrates were performed in a different dilution refrigerator with a similar configuration. Chips were attached to a thermally anchored printed circuit board on the mixing chamber plate of a dilution refrigerator with a base temperature of 10 mK. The devices are shielded from magnetic fields with superconducting and cryoperm cans with a blackbody absorber to suppress stray infrared radiation. The input signal chain has a total attenuation of 76 dB achieved by a series of attenuators anchored at different temperatures, and a 7.65 GHz low-pass filter. The output signal line is filtered by isolators and amplified by a HEMT at 4K and a series of room-temperature amplifiers. Relaxation times are measured in the dispersive regime continuously on each device over multiple days.

\subsubsection{TEM Sample Preparation}
TEM samples were prepared using a 30 kV focused Ga$^+$ ion beam. In order to protect the surface oxide during the ion milling process, the sample was first coated with 50 nm of carbon. The samples were finely polished to a thickness of roughly 50 nm using 5 kV and 2 kV Ga$^+$ ions in an effort to remove surface damage and amorphization in the regions of interest.

\subsubsection{STEM Data Collection}
EDS data was acquired using a JEOL ARM 200CF ARM S/TEM operated at 200 kV. The camera length is set to 8 cm and the condenser aperture is selected to provide a convergence semi-angle of 30 mrad with beam current $\sim$ 20 pA in order to minimize beam-induced damage. The intensity of the Nb L$\alpha$ and Ta L$\alpha$ signals were plotted as a function of position for the Ta-capped Nb sample. The intensity of the Nb L$\alpha$ and Al K$\alpha$ signals were plotted as a function of position for the Al-capped Nb sample. The intensity of the Nb L$\alpha$, Ti K$\alpha$, and N K$\alpha$ signals were plotted as a function of position for the TiN-capped Nb sample. The intensity of the Nb K$\alpha$ and Au L$\alpha$ signals were plotted as a function of position for the Au-capped Nb sample. In order to prepare the chemical maps presented in Fig. 2a-c, the individual chemical maps for each sample were normalized, overlaid, and low-pass filtered using a Gaussian function with $\sigma$ = 2. The EELS data and associated HAADF STEM image was acquired using a Titan Themis with GIF quantum ER system. Electron diffraction patterns from the Ta oxide was collected on JEOL 300F Grand ARM S/TEM using an accelerating voltage of 300kV. In this case, the camera length is set to 20 cm and the condenser aperture is selected to provide a convergence semi-angle of 1 mrad. Four-dimensional STEM (4D-STEM) data sets~\cite{RN33} were acquired in a 200 x 100 mesh (with pixel size of 1.6 nm) across the thin films using a Gatan OneView camera and synchronized using STEMx. 

\clearpage

\subsection{Qubit T1 Measurements}

\begin{table}[h]
\begin{tabular}{|c|c|c|c|c|}

\multicolumn{5}{c}{} \\
\hline
\textbf{Geometry} & \textbf{Film} &  \textbf{Qubit Frequency} (MHz)& \textbf{Average T1 ($\mu$s)} &\textbf{Q} (millions)\\
\hline
A& Nb& 5986&  54&2.0
\\
\hline
A& Nb& 5796&  21&0.8
\\
\hline
A& Nb/TiN& 4990&  60&1.9
\\
\hline
 A& Nb/TiN& 4721&105&3.1
\\
\hline
 A& Nb/TiN& 4645&55&1.6
\\
\hline
 A& Nb/Al& 4894&77&2.4
\\
\hline
 A& Nb/Al& 4968&69&2.2
\\
\hline
 A& Nb/Al& 4586&86&2.5
\\
\hline
 A& Nb/Al& 4512&78&2.2
\\
\hline
 A& Nb/Au& 4637&80&2.3
\\
\hline
 A& Nb/Au& 4077&134&3.4
\\
\hline
 A& Nb/Au& 4185&100&2.6
\\
\hline
 A& Nb/Au& 4193&120&3.2
\\
\hline
 A& Nb/Au& 4657&109&3.2
\\
\hline
 A& Nb/Au& 4749&137&4.1
\\
\hline
 A& Nb/Au& 4782&27&0.8
\\
\hline
 A& Nb/Ta& 5005&102&3.2
\\
\hline
 A& Nb/Ta& 4746&118&3.5
\\
\hline
 A& Nb/Ta& 4707&103&3.0
\\
\hline
 A& Nb/Ta& 4488&161&4.5
\\
\hline
\end{tabular}
\caption{\label{tab:qubits_capping} Summary of qubit devices prepared with different capping layers.}
\end{table}

\begin{table}[h]
\begin{tabular}{|c|c|c|c|c|}

\multicolumn{5}{c}{} \\
\hline
\textbf{Geometry} & \textbf{Film} &  \textbf{Qubit Frequency} (MHz)& \textbf{Average T1 ($\mu$s)} &\textbf{Q} (millions)\\
\hline
 A& Nb/Ta& 3476&131&2.9
\\
\hline
 A& Nb/Ta& 3485&136&3.0
\\
\hline
 A& Nb/Ta& 3477&186&4.1
\\
\hline
 A& Nb/Ta& 3530&120&2.7
\\
\hline
 B& Nb/Ta& 3945&280&6.9
\\
\hline
 B& Nb/Ta& 3952&324&8.0
\\
\hline
 B& Nb/Ta& 3814&262&6.3
\\
\hline
 B& Nb/Ta& 3930&214&5.3
\\
\hline
 C& Nb/Ta& 3737&193&4.5
\\
\hline
 C& Nb/Ta& 3634&199&4.5
\\
\hline
 C& Nb/Ta& 3904&109&2.7\\
\hline
\end{tabular}
\caption{\label{tab:qubit_geometries} Summary of Nb/Ta qubit devices with different geometries.}
\end{table}

\clearpage

\begin{figure}[t]
  \begin{center}
    \includegraphics[width= \columnwidth]{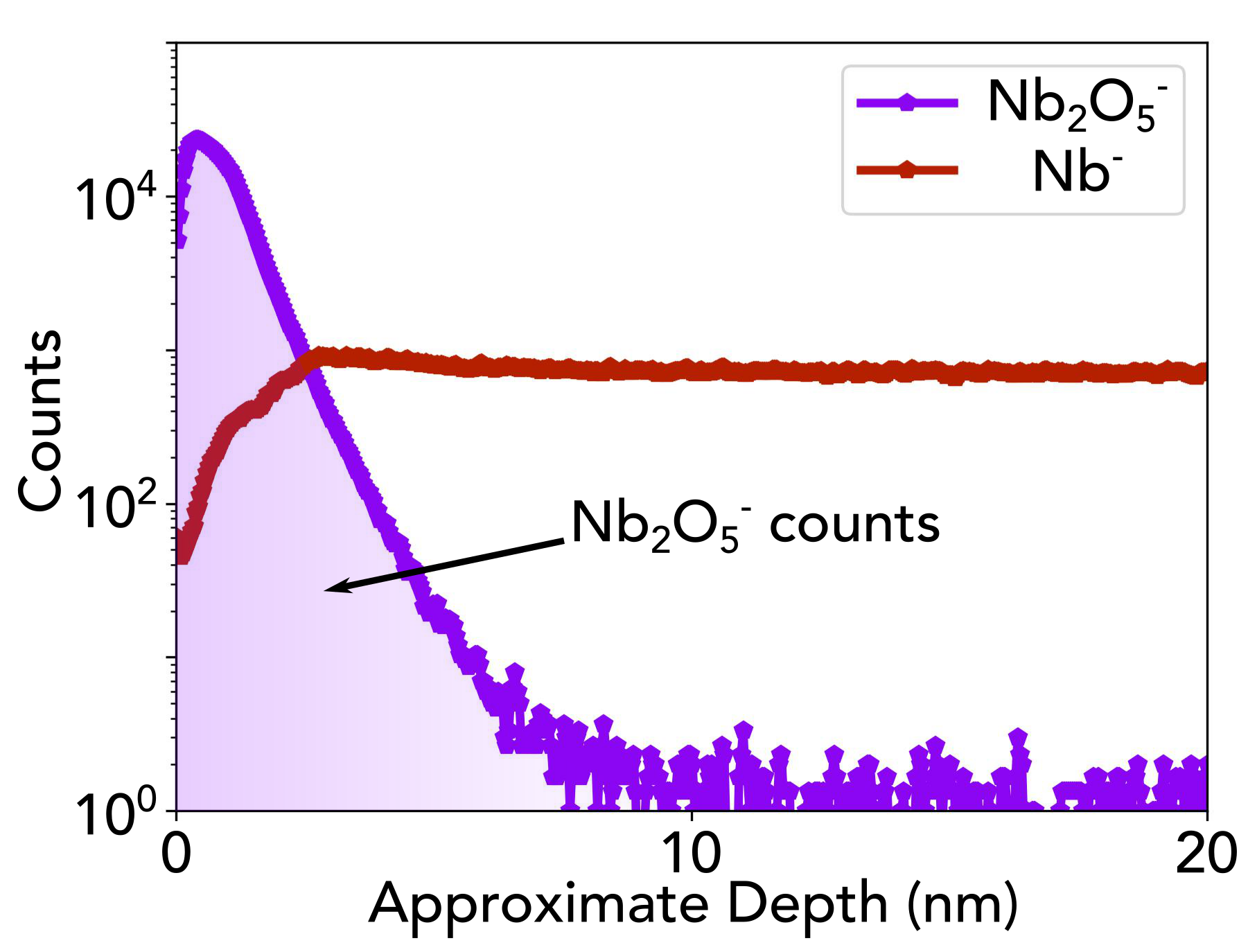}

    \caption{ToF-SIMS Depth profile taken from the surface of Nb film. The Nb$_2$O$_5$ region is labeled. Fig. \ref{fig:characterization}e is constructed by taking the integrated counts of the Nb$_2$O$_5$$^-$ for each of the samples using the same ion beam conditions.}
  	\label{SIMS}
  \end{center}
\end{figure} 

\begin{figure*}[t]
  \begin{center}
    \includegraphics[width= 7in]{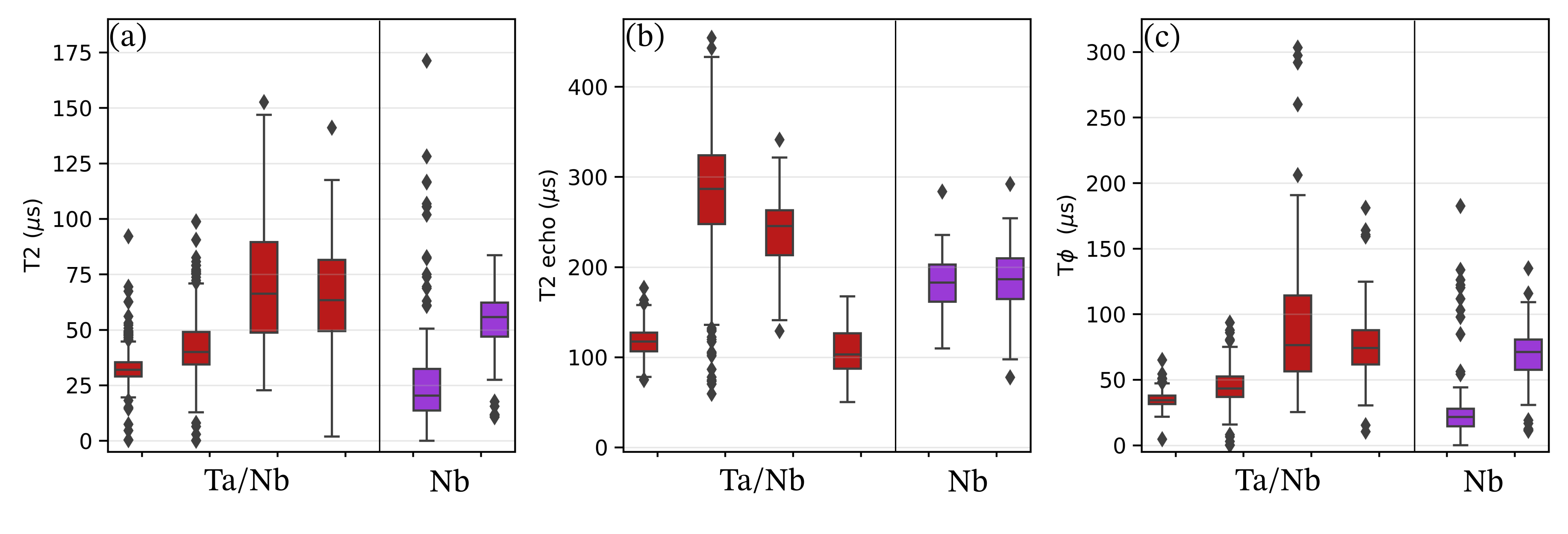}

    \caption{Box plot displaying measured (a) T\textsubscript{2}, (b) T\textsubscript{2} echo, and (c) T$_{\phi}$ values for test devices fabricated on silicon substrates. Boxes mark the 25th percentile and the 75th percentile of the measurement distribution. The line inside each box represents the median value, and whiskers or diamonds represent outliers. The T\textsubscript{2}, T\textsubscript{2} echo, and T$_{\phi}$ values are not heavily impacted by capping Nb with Ta.}
  	\label{T2}
  \end{center}
\end{figure*} 

\clearpage

\begin{figure*}[t]
  \begin{center}
    \includegraphics[width= 7in]{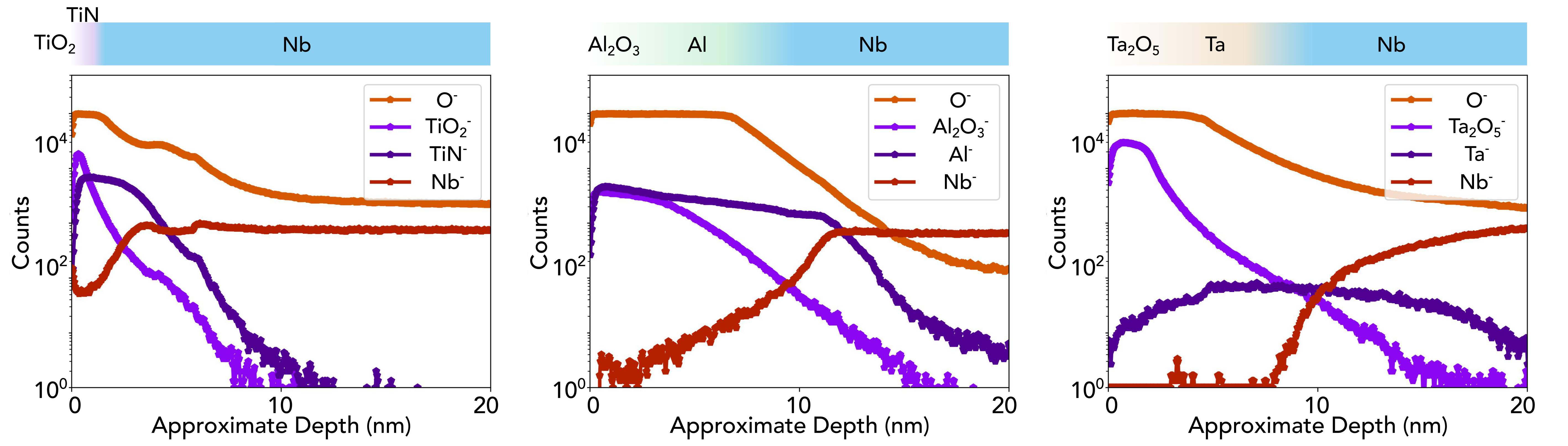}
    \caption{ToF-SIMS depth profiles taken from the surface of Nb films capped with TiN, Al, and Ta. TiO$_2$, Al$_2$O$_3$, and Ta$_2$O$_5$, respectively are observed at the surface of these films.}
  	\label{amorphous_oxides}
  \end{center}
\end{figure*}

\begin{figure*}[t]
  \begin{center}
    \includegraphics[width= 7in]{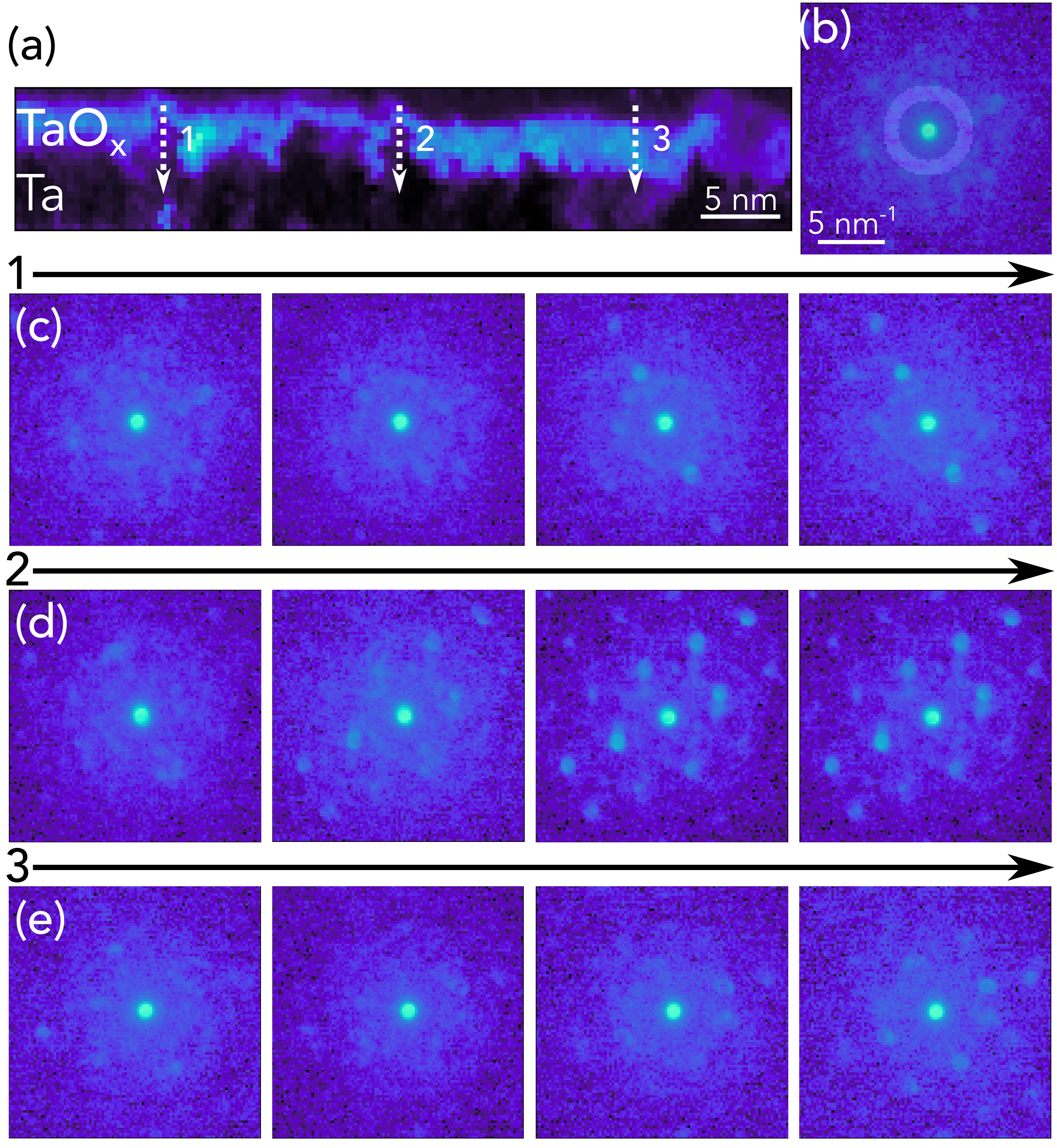}
        \caption{(a) DF image of Ta oxide/Ta interface using a virtual detector that matches the diffraction ring of the Ta oxide. The dotted arrows indicate regions over which changes in the diffraction pattern are provided in (c-e). (b) Diffraction pattern taken from TaO$_x$ with a virtual annular detector. The radius of the inner ring is 2.07 nm$^{-1}$ and the radius of the outer ring is 3.25 nm$^{-1}$ in order to preferentially produce a TaO$_x$ dark field image. (c-f) Changes in the electron diffraction pattern in the direction of the arrows indicated in (a) for the labeled regions 1, 2, 3, and 4, respectively. In all cases, a diffuse diffraction pattern, which is characteristic of an amorphous solid is observed for the TaO$_x$. Intense diffraction spots tend to appear when moving towards the crystalline Ta metal.}
\label{amorphous_oxides_STEM}
\end{center}
\end{figure*} 

\begin{figure*}[t]
\includegraphics[width=7in]{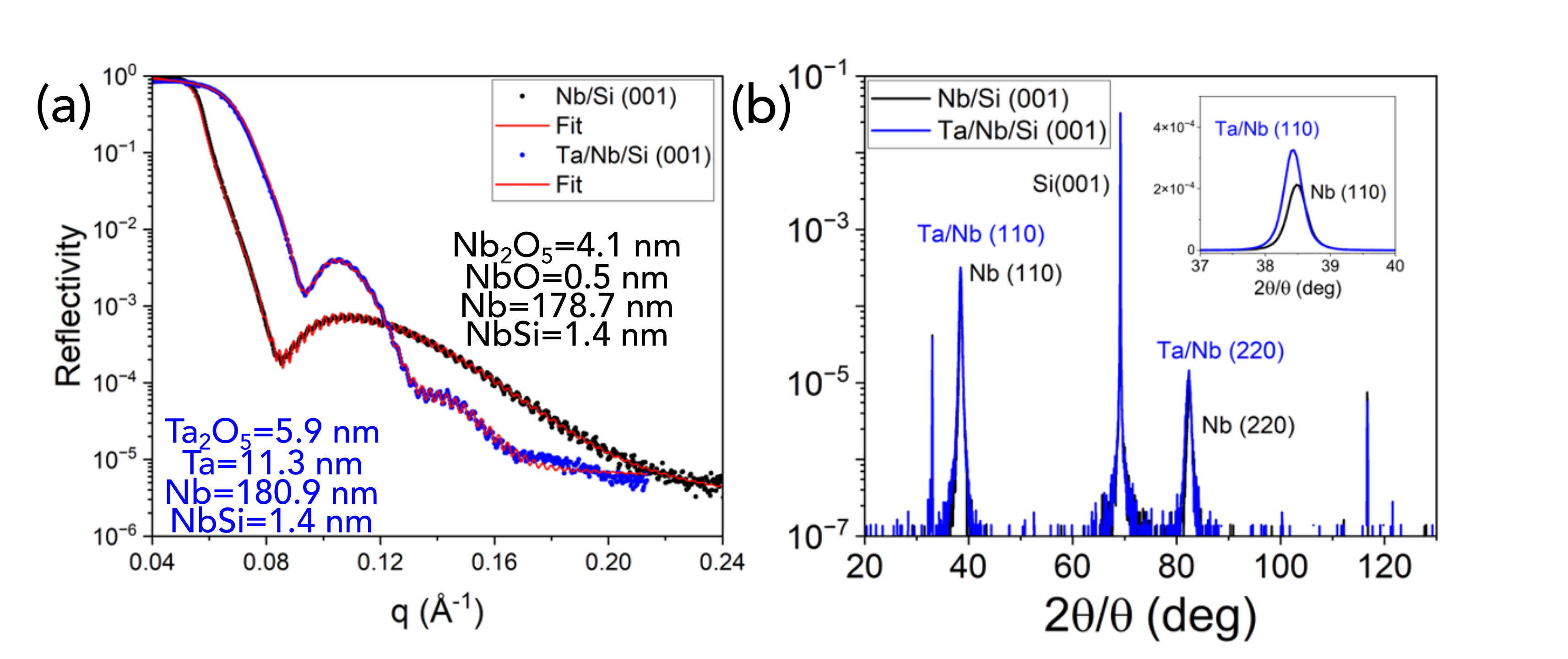}
\caption{X-ray reflectivity (a) and x-ray diffraction (b) pattern taken from both the Nb sample and the Nb sample capped with Ta. Fits based on dynamical scattering suggests that the surface oxide of Nb consists of roughly 4.1 nm of Nb$_2$O$_5$ and 0.5 nm of NbO whereas the surface oxide of Ta consists entirely of 5.9 nm of Ta$_2$O$_5$. X-ray diffraction pattern suggest both sets of films exhibit the body-centered cubic (BCC) crystal structure and predominantly exhibit \{110\} texture. For Ta, this crystal structure is associated with the $\alpha$ phase.}
\label{fig:xrr}
\end{figure*}

\end{document}